\documentclass[twocolumn,preprintnumbers,amsmath,nofootinbib, amssymb]{revtex4}
\usepackage{graphicx}
\usepackage{epstopdf}

\usepackage{graphicx}
\usepackage{bm}
\usepackage{graphicx}

\usepackage{psfrag}  
\usepackage{amsmath,amsthm}
\usepackage{amsfonts}
\usepackage{amssymb}
\usepackage{bm}

\newcounter{x}
\newtheorem{inneraxiom}{Axiom}
\newenvironment{axiom}[1]
  {\renewcommand\theinneraxiom{\arabic{x} (#1)}\inneraxiom\stepcounter{x}}
  {\endinneraxiom}

\begin{document}

\title{Zipf's law, unbounded complexity and open-ended evolution}

\author{Bernat Corominas-Murtra$^{1}$, Lu\'is F Seoane$^{2,3,4}$ and Ricard V. Sol\'e$^{3,4,5}$}

\address{ 
			(1) Institute of Science and Technology Austria, Am Campus 1, A-3400, Klosterneuburg, Austria\\
			(2) Department of Physics, Massachusetts Institute of Technology, 77 Massachusetts Ave, Cambridge, MA 02139, USA\\
			(3) ICREA-Complex Systems Lab, UPF-PRBB. Dr Aiguader 88, 08003 Barcelona, Spain\\ 
			(4) Institute Evolutionary Biology, UPF-CSIC, Pg Maritim Barceloneta 37, 08003 Barcelona\\
			(5) Santa Fe Institute, 1399 Hyde Park Road, 87501 Santa Fe, New Mexico, USA
		 }
	  
\begin{abstract}

  A major problem for evolutionary theory is understanding the so called {\em open-ended} nature of evolutionary change,
from its definition to its origins. Open-ended evolution (OEE) refers to the unbounded increase in complexity that seems
to characterise evolution on multiple scales. This property seems to be a characteristic feature of biological and
technological evolution and is strongly tied to the generative potential associated with combinatorics, which allows the
system to grow and expand their  available state spaces. 
Interestingly, many complex systems presumably displaying OEE, from language to proteins,
share a common statistical property: the presence of Zipf's law. Given an inventory of basic items (such as words or
protein domains) required to build more complex structures (sentences or proteins) Zipf's law tells us that most of
these elements are rare whereas a few of them are extremely common. Using Algorithmic Information Theory, in this paper
we provide a fundamental definition for open-endedness, which can be understood as {\em postulates}. Its statistical
counterpart, based on standard Shannon Information theory, has the structure of a variational problem which is shown to
lead to Zipf's law as the expected consequence of an evolutionary process displaying OEE. We further explore the
problem of information conservation through an OEE process and we conclude that statistical information (standard
Shannon information) is not conserved, resulting into the paradoxical situation in which the increase of information
content has the effect of erasing itself. We prove that this paradox is solved if we consider non-statistical forms of
information. This last result implies that standard information theory may not be a suitable theoretical framework to
explore the persistence and increase of the information content in OEE systems.

\end{abstract}

\keywords{Complexity, algorithmic complexity, open-ended evolution, Zipf's law} 


	\thanks{bernat.corominas-murtra@ist.ac.at, ricard.sole@upf.edu}
\maketitle


\section{Introduction}
	\label{sec:1}

	Life has been evolving on our planet over billions of years, undergoing several major transitions along with multiple
	events of both slow and rapid change affecting structure and function \cite{Eors, SchusterEVolComplex, LifeAscending,
	MST1}. Life seems to be indefinitely capable of increasing in complexity. This is illustrated, as an instance, by the
	trend towards larger genomes and diverse cell types exhibited by multicellular organisms. Moreover, the emergence of
	high neuronal plasticity and complex communication provided the substrate for non-genetic modes of adaptation. A key
	concept that pervades many of these innovations is the idea that evolution is ``open-ended''. Following
	\cite{KepaOEE}, Open-Ended Evolution (OEE) can be defined as follows: ``a process in which there is the possibility
	for an indefinite increase in complexity.'' What kind of systems can exhibit such unbounded growth in complexity
	\cite{OpenProblemsAlife}? What are the conditions under which the complexity --and thus, the information content of
	the system-- can increase and what are the footprints of such an open-ended increase of complexity? Which kind of
	information is encoded in an OEE system? The aim of this paper is to give hints to the these questions. 

	Open-ended evolutionary change needs a dynamical behaviour allowing complexity to grow  in an unbounded way
	\cite{KepaOEE, Nehaniv2000}. This requires a very large exploration space but this is only a necessary requirement. For example, as
	noticed in \cite{Day2012} mathematical models used in population genetics involving infinite alleles --using Markov
	models-- do not display open-ended evolution. 
	Previous attempts to address the problem of OEE involved different approximations and degrees of abstraction. John von
	Neumann was one of the early contributors to this issue \cite{vonNeumann1949, McMullin2000, KepaOEE}. In all these
	studies, some underlying mechanism is assumed to be operating, and arguments are made concerning the presence of
	self-replication, genotype-phenotype mappings, special classes of material substrates and physico-chemical processes
	\cite{KepaJuli, KepaOEE}. On the other hand, a theory of OEE might demand a revision of the role of novel niches and
	abiotic changes, as well as refining what we understand as  the open-endedness of a system \cite{TaylorWiser2016,
	deVladarSzathmary2017}. Special suitable candidates for OEE systems are
	complex systems exhibiting generative rules and recursion. The best known case is human language. Thanks to recursion, syntactic rules
	are able to produce infinite well-formed structures and thereby the number of potential sentences in
	a given language is unbounded \cite{Hauser2002}. In another example, Darwinian evolution proceeds through tinkering
	\cite{Jacob1977,Sole2002}, continuously reusing existing parts. These are first copied --hence bringing in some
	redundancy into evolving systems-- but are later on modified through mutation or recombination. Despite the obvious
	differences existing between Darwinism in biology and human-guided engineering \cite{Jacob1977}, this process of
	tinkering appears to be common too in the growth of technological systems, thus indicating that copy-and-paste
	dynamics might be more fundamental than expected \cite{Sole2013}. 

	\begin{figure*}
		\begin{center} 
			\includegraphics[width=13.5cm]{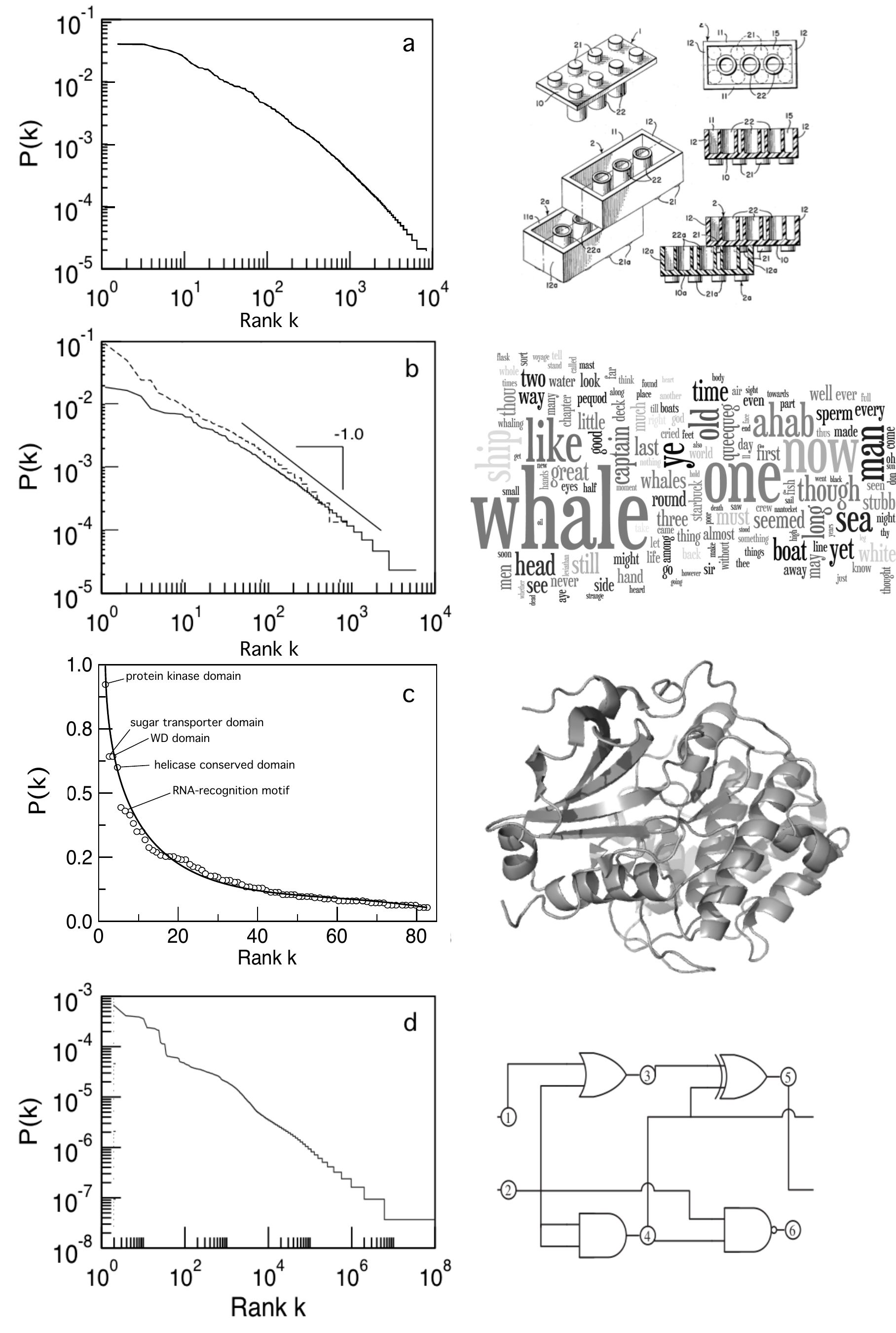}
		\end{center} 

    \caption{Zipf's law distributions are commonly found in very different systems candidates to display open-endedness.
Here we show several examples of scaling behavior involving (a) Lego systems, (b) written language and (c) proteins. In
(a) we display (in log scale) the probability of finding the $i$-th most abundant type of LEGO brick within a very large
number of systems (see details in \cite{LEGOzipf}). In (b) the log-scale rank-size distribution of Herman Melville's
Moby Dick is displayed. The dashed line shows the frequency versus rank for words having length 5, which is the average
length of words in this particular book. The linear rank plot displayed in (c) shows the corresponding rank distribution
of protein folds in a large protein database (redrawn from \cite{Searls2002}). The line is a power law fit. Here the
names of some of the domains, which are associated to particular functional traits, are indicated. (d) Zipf's law in the
frequency of logic modules used in evolved complex circuits (adapted from \cite{Raman2011}).}
		\label{fig:01}   
	\end{figure*}

	These systems are very different in their constitutive components, dynamics, and scale. However, all share the
	presence of a common statistical pattern linked to their diversity: fat tailed distributions. Four examples are
	provided in figure (\ref{fig:01}). In all these cases, the frequency distribution of the basic units decays following approximately
	Zipf's law. Zipf's law was first reported for the distribution of city sizes \cite{Auerbach:1913}, and then
	popularised as a prominent statistical regularity widespread across all human languages: in a huge range of the vocabulary, the frequency of any word is
	inversely proportional to its rank \cite{Zipf1949,NewmanZipf}. Specifically, if we rank all the occurrences of words
	in a text from the most common word to the less common one, Zipf's law states that the probability $p(s_i)$ that in a
	random trial we find the $i$-th most common word $s_i$ (with $i=1,...,n$) falls off as
		\begin{eqnarray}
			p(s_i) &=& \frac{1}{Z}i^{-\gamma}\quad,
			\label{PLAnsatz}
		\end{eqnarray} 
	with $\gamma \approx 1$ and $Z$ the normalization constant, i.e., $Z=\sum_{i\leq n}i^{-\gamma}$. Stated otherwise, the
	most frequent word will appear twice as often as the second most frequent word, three times as often as the third one,
	and so on. This pattern is found in many different contexts and can emerge under different types of dynamical rules
	--see \cite{NewmanZipf,ZipfUniversal, CorominasMurtraSole2011, Corominas-Murtra:2015, Corominas-Murtra:2016} and
	references therein.

	The examples shown in figure (\ref{fig:01}) involve: (a) LEGO$^{\circledR}$ models, (b) human language, (c) proteins
	and (d) evolved electronic circuits. The first example provides an illustration of structures emerging through 
	copy-paste and combination in a non-biological setting. This toy system allows to exploit the intrinsic combinatorial
	explosion associated to the multiple ways in which different bricks can be interlinked. In figure (\ref{fig:01}a) we
	plot the number of times that each type of brick occurred within a very large data set of Lego models \cite{LEGOzipf}.
	The rank plot reveals that some simple bricks --as those shown in figure (\ref{fig:01}a), right-- are extremely common
	whereas most bricks, having more complex shapes and larger size, are rare. The analysis showed that the statistical
	distribution can be well fitted using a generalized form of equation \ref{PLAnsatz} known as the Pareto-Zipf
	distribution. This reads:

		\begin{eqnarray}
			p(s_i) &=& \frac{1}{Z}(i+i_0)^{-\gamma}\quad, 
			\label{Paretolaw}
		\end{eqnarray}
	where $Z$ is again the corresponding normalization and $i_0$ a new parameter that allows to take into account the
	curvature for small $i$-values. This picture is similar to the one reported from the study of large written corpora,
	as illustrated in figure (\ref{fig:01}b) \cite{ACSzipf}.
	Our third example is given by so called protein domains, which are considered the building blocks of protein
	organization and an essential ingredient to understand the large scale evolution of biological complexity
	\cite{Koonin2002, Teichmann2004, Kim2012,Carlos2005}. Here each protein domain --or {\em fold}-- is characterized by its
	essentially independent potential for folding in a stable way and each protein can be understood as a combination of
	one, two, or more domains. In figure (\ref{fig:01}c) the rank distribution of observed folds from a large protein
	database is displayed. Domains define the combinatorial fabric of the protein universe and their number, although
	finite, has been increasing through evolution \cite{Kim2012}. The fourth example gives the frequency of use of
	4-element modules within complex circuits \cite{Raman2011}.

	The repertoire of LEGO$^{\circledR}$ bricks, words, protein domains, and circuit modules provide the raw materials to
	combinatorial construction; but they also share the underlying presence of a grammar, to be understood here as the
	compact description of a language. As indicated in \cite{LEGOzipf}, if we treat pieces of LEGO$^{\circledR}$ as words
	and models as utterances, LEGO$^{\circledR}$ appears as a class of artificial language and the resulting structures
	are passed from generation to generation through cultural transmission. This is of course a largely metaphoric
	picture, since the final outcome of the combinatorics is usually a non-functional design, unbounded by the potential
	combinations but not by functional constraints. This might actually be the reason why its statistical distribution,
	described by equation (\ref{Paretolaw}) deviates from equation (\ref{PLAnsatz}). Protein domains too exhibit a grammar
	in which a set of generative rules for combining the available folds provides an explanatory mechanism for the
	observed repertoire of protein structures \cite{Searls2002,Rose2002, Gimona2006}. In summary, these systems --and
	others like electronic circuits or genomes, molecular networks \cite{neutral1, neutral2, neutral3} and complex
	circuits \cite{neutralcircuits1} and even evolved technology \cite{Arthur2006}-- are characterized by a growth process
	that is expanding their inventories over time, the presence of generative rules allowing new structures to emerge, and
	a common statistical pattern described by Zipf's law.

	In this paper we provide a general definition, or {\em postulates} of OEE based on Algorithmic Information Theory, and
	we show that the common presence of Zipf's law in these seemingly disparate systems may be deeply connected to their
	potentially open-ended nature. Furthermore, we explore the consequences that OEE has for the conservation of the
	information, identifying the {\em information loss paradox} in OEE systems. This paradoxical situation, in which the
	system loses all its past information in the long run, even though the step-by-step information transmission is
	maximized, is shown to be a problem of the statistical nature of Shannon Information Theory. Indeed, we prove that, in
	the general setting of Algorithmic Information Theory, information can be conserved and systems can grow without
	bounds without removing the traces of its past. Therefore, the general study of OEE systems must be framed in a
	theoretical construct not based on standard information theory, but in a much more general one, inspired in 
	non-statistical forms of information content. We finally observe that the connection of fundamental results of
	computation theory, and even G\"odel's incompleteness theorem, with general problems of evolutionary theory has been
	approached before in \cite{Adami1998,Day2012, HernandezZenil2018}.
 
\section{Algorithmic information theory}
	\label{sec:2}
		
	Algorithmic Information Theory (AIT) \cite{Solomonoff1960, Chaitin1977, Chaitin1990, KCkolmogorov, KCsolomonoff,
	KCchaitin, AdamiCerf, Zurek1989, ChaitinBooks, Li:1997, Cover1991} is a natural framework to address the problem of
	OEE. It incorporates powerful (still unexplored) tools to model the complexity of living systems, which, for example,
	has often been associated to information storage in the genome \cite{Adami1998, AdamiCerf}. Such kind of information
	results from the growth of genome complexity through both gene duplication and the interactions with the external
	world and is (by definition) a path-dependent process. Here we consider that we {\em encode} our evolving system into
	strings of symbols. We assume that, as long as the system evolves, such descriptions can grow and change, in a 
	path-dependent way. As we shall see, the derived abstract framework is completely general, and applies to any system
	susceptible of displaying OEE.

	A natural question arises when adopting such an abstract framework: Why using Kolmogorov Complexity for our approach
	to open-ended evolution? The first reason is that it is based on strings obtained from a given alphabet, which
	naturally connects with a representation based on sequences \cite{AdamiCerf} such as those in some of our examples
	from figure (\ref{fig:01}). Second, it connects with information theory (which is the most suitable coarse-grained
	first-approximation to biology \cite{Hopfield1994}) resulting on a more fundamental framework. Third, it consistently
	distinguishes in a meaningful way predictable from unpredictable sequences, and how these scale with size. Finally,
	the algorithmic definition based on the use of a {\em program} matches our intuition that evolution can be captured by
	some computational picture.

\begin{figure}
			\begin{center} 
				\includegraphics[width = 8.5 cm]{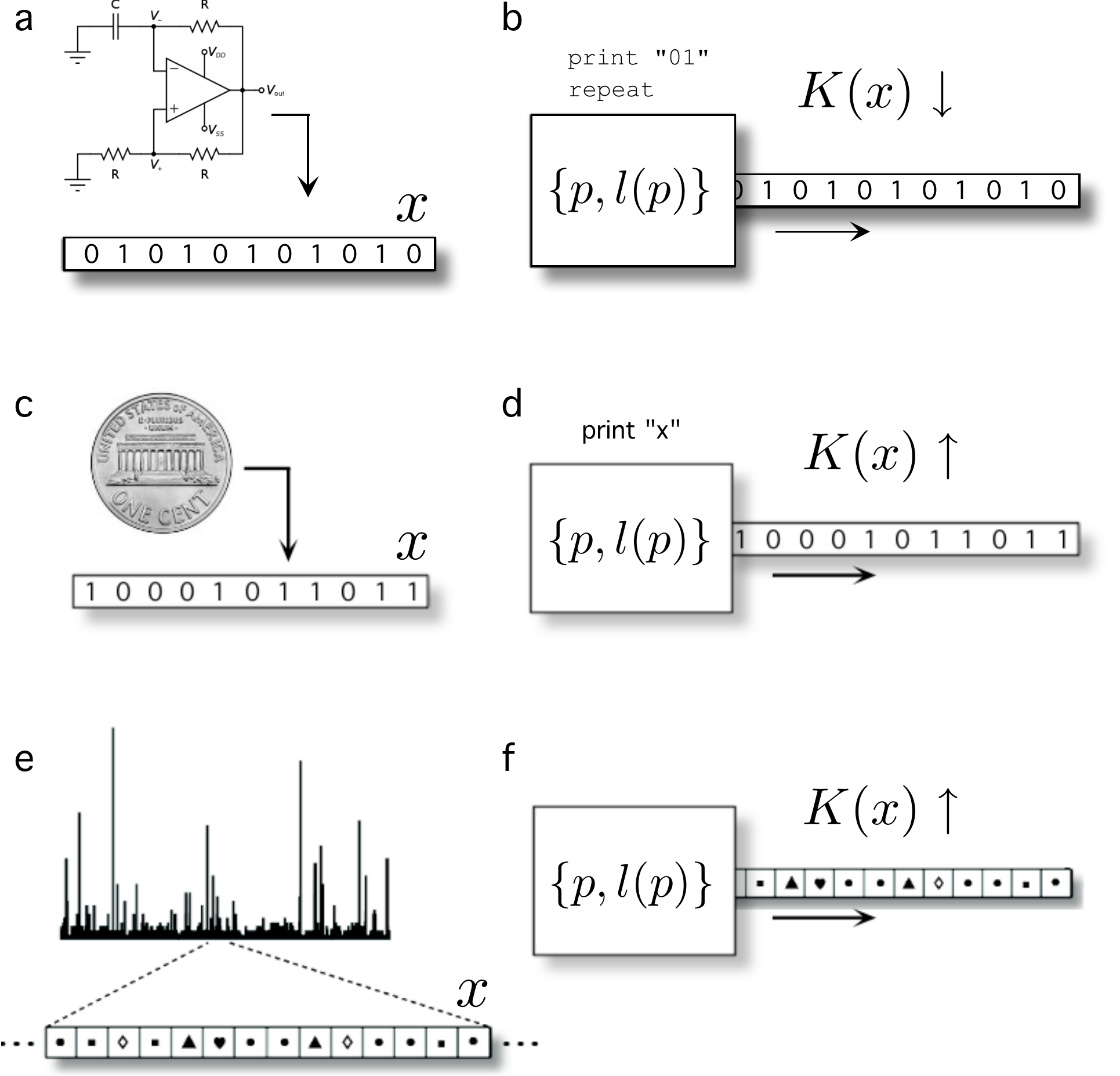}
			\end{center} 
      \caption{Measuring string complexity. If $\ell(p)$ is the length of the program, the Kolmogorov complexity $K(x)$
is defined as the smallest program able to write $x$. Simple dynamical systems (a) such as oscillators produce
predictable, simple strings (b) thus having a low complexity. On the other extreme (c) a coin toss creates a completely
random sequence and the program (d) is such that $K(x)=\ell(x)$. A system exhibiting broad distributions (e-f) and a
large set of states is also likely to display high $K(x)$.}
			\label{fig:02}
		\end{figure}

		Let us first introduce a key concept required for our analysis: Kolmogorov --or algorithmic-- complexity,
		independently developed by Kolmogorov \cite{KCkolmogorov}, Somolonoff \cite{KCsolomonoff}, and Chaitin
		\cite{KCchaitin}. Roughly speaking, if a given process can be described in terms of a string of bits, the complexity
		of this string can be measured as the shortest computer program capable of generating it \cite{Zurek1989,
		ChaitinBooks}. The underlying intuition behind this picture --see figure (\ref{fig:02})-- is that {\em simple},
		predictable strings, such as $10101010101010...$ can be easily obtained from a small piece of code that essentially
		says `write ``10''' followed by `repeat' as many times as needed. This would correspond to a regular system, such as
		a pendulum or an electronic oscillator --see figure (\ref{fig:02}a-b)-- and the simple dynamical pattern is
		reproduced by a short program. Instead, a random string generated by means of a coin toss (say
		$0100110011101101011010...$) would only be reproduced by using a program that writes exactly that sequence and is
		thus as long as the string itself --figure (\ref{fig:02}c-d). Other stochastic processes generating fluctuations
		--figure (\ref{fig:02}e-f)-- and represented as strings of $n$ bits can be similarly described, and their complexity
		shall lie somewhere between both extremes.

		The stochasticity inherent to the most algorithmically complex strings (e.g. a coin toss, as introduced above)
		invites us to think in terms of {\em statistical} or {\em information entropy}. But the Kolmogorov complexity is,
		conceptually, a more fundamental measure of the complexity of such processes \cite{Li:1997, Cover1991}. A formal
		definition follows. Let $x$ and $p$ be finite binary strings of length $\ell(x)$ and $\ell(p)$ respectively. Let
		${\cal T}_u$ be a universal Turing machine. Note that a finite binary string $p$ can define the computations that a
		universal Turing machine \cite{TuringOnComputable}  will implement when $p$ is fed as an input -- i.e. it can define
		{\em programs} executed by the Turing machine. We will consider a set of {\em prefix free} programs: In such a set
		of programs, no program is the prefix of another program. This property is crucial for most of the results of AIT or
		even standard information theory \cite{Li:1997, Cover1991}.  Let ${\cal T}_u(p)$ denote the output of the computer
		${\cal T}_u$ when running the program $p$. Considering now all possible programs $p$ that produce $x$ as an output
		when fed into ${\cal T}_u$, the (prefix free) Kolmogorov Complexity $K_{{\cal T}_u}(x)$ of the string $x$ with
		respect to the universal computer ${\cal T}_u$ is defined as \cite{Li:1997}:
			\begin{eqnarray}
				K_{{\cal T}_u}(x) &=& \min_{p:{\cal T}_u(p)=x}\left\{\ell(p)\right\}\quad.
				\label{eq:KolmogorovComplexity}
			\end{eqnarray}
		This quantity is computer independent up to an additive constant \cite{Li:1997,Cover1991} so we will omit the
		subindex when referring to it. If $x$ is a random string, we would have a simple relation:
			\begin{eqnarray}
				K(x) &=& \ell(x)\quad,
				\label{eq:KolmogorovComplexity2}
			\end{eqnarray}
		since all $\ell(x)$ bits need to be included, and we say that the sequence $x$ is {\em incompressible}.

		In addition, and as it happens with the statistical entropy, one can define the {\em conditional} algorithmic
		complexity as follows: Let $x$, $y$, and $p$ be finite binary strings again and let ${\cal T}^y_u$ be a universal
		Turing machine to which a description of $y$ has already been made available. The Kolmogorov complexity of $x$ given
		$y$ is the length of the shortest program $p$ that, when applied to a universal Turing machine, modifies $y$ to
		display $x$ as an output:
			\begin{eqnarray}
				K(x|y)=\min_{p:{\cal T}^y_u(p)=x}\left\{\ell(p)\right\}\quad .
				\label{eq:CondKolmogorovComplexity}
			\end{eqnarray}
		Notice that even though $K(x)$ can be arbitrarily large, $K(x|y)$ accounts for the minimum program that knows the
		differences between $x$ and $y$ and amends them\footnote{This quantity has been used as a conditional  complexity
		within the context of evolved symbolic sequences \cite{AdamiCerf}. In this case, $K(s|e)$  referred to the length of
		the smallest program that gives the string $s$ from a given environment $e$,  also defined as a string.}.

	\section{General conditions for Open-Ended Evolution: Postulates}
		\label{sec:postulates}

		We shall concern ourselves with dynamic systems whose description can be made in terms of finite binary strings
		$\sigma_t$ at each time step $t$ over evolutionary time. The {\em complexity} of such an object at time $t$ is given
		by $K(\sigma_t)$. This object shall evolve through intermediate steps in a path-dependent manner; thus the
		quantities $K(\sigma_t)$, $K(\sigma_{t+\Delta t})$, and $K(\sigma_{t+\Delta t}|\sigma_t)$ and the relationships
		between them will play a paramount role.

		Let $\sigma_t$ be the description of the system at time $t$. Let the sequence $\Sigma(t) \equiv \{\sigma_1,
		\sigma_2,\dots, \sigma_t\}$ be the {\em history} of the system until time $t$ in arbitrary time units. We want the
		process that builds $\sigma_t$ to be an open-ended evolutionary one, hence we turn our attention to the complexity
		of its evolutionary history $\Sigma(t)$. A minimal condition that this historical process has to obey to be called
		open-ended is that its complexity (properly normalized) always increases:
			\begin{axiom}{Open-Endedness}
				We say that the process that generates $\sigma_t$ is open-ended if 
					\begin{eqnarray}
						{K(\Sigma(t)) \over t} &\leq& {K(\Sigma(t+1)) \over t+1}\quad,
						\label{eq:K<K1}
					\end{eqnarray}
				for all $t=1, \dots, \infty$. 
				\label{ax:01}
			\end{axiom}
		Of all open-ended processes that obey equation (\ref{eq:K<K1}), we are interested in those whose complexity is not
		bounded:
			\begin{axiom}{Unboundedness}
				We say that the process generating $\sigma_t$ has an unbounded complexity if for any natural number
				$N\in\mathbb{N}$ there is a time $t$ such that
					\begin{eqnarray}
						{K(\Sigma(t)) \over t} &>& N \quad.
						\label{eq:unbounded1}
					\end{eqnarray}
				\label{ax:02}
			\end{axiom}
		These two axioms imply that information is always being added by the generative process in the long term -- hence
		more bits are needed to describe later stages of the evolutionary history. The knowledge of the history up to time
		$t$ is not enough to predict what will happen next. If it were, the description of later stages of the evolutionary
		history would be implicit in the description of the history at time $t$, and Axiom \ref{ax:01} would be violated.
		Equation (\ref{eq:unbounded1}) also implies that the information of the processes we are interested in will never
		converge, eventually diverging for large times. These equations do not impose any condition on the complexity of the
		system at a given time step. Notably, i) they admit a situation in which the description of the system -- but not of
		its history -- drops ($K(\sigma_t) > K(\sigma_{t+1})$, which might happen in biology \cite{NeefMoya2011}, see also
		figure S1) and ii) they do not imply any connection between states $\sigma_t$ and $\sigma_{t+1}$. This second point
		is possible because we have not imposed yet that this is an evolutionary process. We would hardly call a process
		`evolutionary' if its successive steps are completely unrelated, hence:
			\begin{axiom}{Heredity principle}
				Evolutionary processes attempt to minimize the action 
					\begin{eqnarray}
						S\left(\Sigma(t) \rightarrow \Sigma(t+1) \right) \equiv K\left(\Sigma(t+1) | \Sigma(t)\right) \quad. 
						\label{eq:heredity}
					\end{eqnarray}
				\label{ax:03}
			\end{axiom}

		That is, evolutionary processes try to minimize the amount of operations implemented to move the system from one
		state to the next, under whichever other constraints might apply. In the case of open-ended evolutionary systems,
		they try to minimize the number of operations needed to unfold in time while always increasing the informational
		content of the evolutionary history (as equations (\ref{eq:K<K1}) and (\ref{eq:unbounded1}) demand). We could apply
		the same axiom, say, to Darwinian evolutionary processes saying that they attempt to minimize equation
		(\ref{eq:heredity}) subjected to random mutation and selection. (Note that this  has no saying on whether Darwinian
		processes are inherently open-ended or not.) Axiom \ref{ax:03} defines an AIT-like least action principle that
		imposes that the information carried between successive steps is maximized as much as other constraints allow, thus
		turning the generative process into a path-dependent one. Without the Heredity principle we could end up with a
		sequence of totally unrelated objects -- i.e. a purely random, unstructured process hardly interpretable as an {\em
		evolving} system.

		We take these axioms as our most general postulates of OEE. In a nutshell, our working definition of open-endedness
		implies that the size of the algorithm describing the history of the system does not converge in time. Therefore,
		even if every evolutionary stage accepts a finite algorithm as a description, the evolutionary path is
		asymptotically uncomputable. These postulates are assumed to be satisfied by all open-ended systems. However, they
		turn out to be too generic to extract conclusions of how OEE systems may behave or which kind of observable
		footprints are expected from them. To gain a greater insight about the effects of OEE we can study a {\em strong}
		version of these postulates that applies not to evolutionary histories, but to objects themselves. Hence we demand
		that:
			\begin{eqnarray}
				K(\sigma_t) \le K(\sigma_{t+1}) \quad,
				\label{eq:strongOE}
			\end{eqnarray}
		at any $t = 1, \dots, \infty$ and that for every natural number $N\in\mathbb{N}$ there is a time $t$ such that 
			\begin{eqnarray}
				K(\sigma_t) > N\quad. 
				\label{eq:strongUnbound}
			\end{eqnarray}
		Also, in the strong version of OEE the action:
			\begin{eqnarray}
				S(\sigma_{t} \rightarrow \sigma_{t+1}) \equiv K(\sigma_{t+1} | \sigma_t) \quad,
				\label{eq:strongHeredity}
			\end{eqnarray}
		is minimized, constrained by equations (\ref{eq:strongOE}) and (\ref{eq:strongUnbound}). As before, $S(\sigma_t
		\rightarrow \sigma_{t+1})$ denotes an informational entropy --or missing information-- when inferring $\sigma_{t+1}$
		from $\sigma_t$. We know from \cite{Zurek1989} that this quantity is bounded by:
			\begin{eqnarray}
				S(\sigma_{t+1}\to \sigma_{t}) \geq |K(\sigma_{t+1}) - K(\sigma_t)|+{\cal O}(1)\quad.
				\label{eq:boundZurek}
			\end{eqnarray}

		As discussed before, the general OEE postulates --equations (\ref{eq:K<K1}-\ref{eq:heredity})-- allow for the
		complexity of $\sigma_t$ to drop, so such processes are not necessarily OEE in the strong sense defined by equations
		(\ref{eq:strongOE}-\ref{eq:strongHeredity}). However, it can be proved that every unbounded OEE process in the
		general sense must contain an unbounded OEE process in the strong sense --see appendix material. That is, the
		strong version of OEE still can teach us something about the footprints of open-ended evolution. 

\section{Statistical Systems: A variational approach to open-ended evolution}
	\label{sec:3}


	We will explore now the consequences of the definition stated above for systems that accept a description --possibly
	partial-- in terms of statistical ensembles. The aim is to  write the three conditions
	for OEE described by equations (\ref{eq:strongOE}-\ref{eq:strongHeredity}) in the language of statistical information
	theory. We will assume now that the statistical properties of this very finite string $\sigma_t$ are themselves
	accurately accounted for by a random variable $X_t$. In other words: We consider that the string $\sigma_t$ as a sequence of observations at the system at time $t$. This will provide a description of the system, in terms of observable states, at time $t$. We further consider that such a description of the system is the outcome of a random variable $X_t$. Then, the algorithmic complexity $K(\sigma_t)$ is of the order of the Shannon entropy $H(X_t)$ associated to the random variable $X_t$ \cite{Li:1997, Cover1991}:
		\begin{eqnarray} 
			K(\sigma_t) & = & H(X_t)+{\cal O}(1)\quad. \nonumber
		\end{eqnarray} 
	Recall that this is the minimal information required to describe the behaviour of a single outcome of $X_t$, not a
	{\em a sequence of trials} of the random variable\footnote{Rigorously speaking, one should say that, if $\sigma$ is
	the description in bits of the outcomes of $N$ trials of $X_t$, then $\frac{K(\sigma)}{N}\to H(X_t)$. } $X_t$. This random variable will represent an observation or realization of the system. Assume
	that we discretise the time, so we use the subscript $n$ or $m$ instead of $t$, and that we label the states $i=1, . .
	., n$.  Now let us define the following family of nested subsets:
		\begin{eqnarray}
			\Omega_1 &=& \{1\} \nonumber \\
			\Omega_2 &=& \Omega_1\cup \{2\}=\{1,2\} \nonumber \\
			\dots &=& \dots \nonumber\\
			\Omega_{n+1} &=& \Omega_n\cup\{n+1\}=\{1,. . .,n+1\} \quad.\nonumber  
		\end{eqnarray}
	The open-ended evolutionary process will traverse the above family of  nested subsets, adding a new state per
	evolutionary time step. We now define a sequence of different random variables
		\[
			X_1, \dots, X_n \quad,
		\]
	such that $X_k$ takes values over the set $\Omega_k$ and follows the probability distribution $p_k(1), . . .,p_k(k)$,
	with $\sum_{i\leq k}p_k(i)=1$. Then:
		\begin{eqnarray}
			H(X_n) = -\sum_{i\leq n}p_n(i)\log p_n(i)\quad. \nonumber
		\end{eqnarray}
	The variational principle derived from the path-dependent process implies now the minimization of the conditional
	entropy of the random variable $X_{n+1}$ given the random variable $X_{n}$, namely:
		\begin{eqnarray}
			H(X_{n+1}|X_n) = -\sum_{i\leq n}p_{n}(i)\sum_{k\leq n+1}\mathbb{P}_n(k|i)\log \mathbb{P}_n(k|i)\;,\nonumber
		\end{eqnarray}
	where $\mathbb{P}_n(k|i)\equiv \mathbb{P}(X_{n+1}=k|X_n=i)$. We will finally assume (without loss of generality) that
	the probability distributions $p_2, . . ., p_{n}$ are sorted in decreasing order,~i.e.:
		\begin{eqnarray}
			p_k(1) > p_k(2) > \dots > p_k(k) \quad.\nonumber
		\end{eqnarray}
	In the appendix  we discuss the conditions under which the consecutive achievement of ordered
	probability distributions is possible.

	Therefore, for statistical systems, the previous constraints for open-endedness from equations (\ref{eq:strongOE}) and
	(\ref{eq:strongUnbound}) must now be rewritten as follows: First,
		\begin{eqnarray}
			H(X_n)  \leq  H(X_{n+1})\quad,
			\label{eq:H<H}
		\end{eqnarray}
	and, for any $N\in\mathbb{N}$, there will be a $n$ such that
		\begin{eqnarray}
			H(X_n)>N\quad.
			\label{eq:unboundedH}
		\end{eqnarray}
	In addition, path dependence condition stated in equation (\ref{eq:strongHeredity}) implies that: 
		\begin{equation}
			{\rm minimize}\;\;H(X_{n+1}|X_n)\quad. 
			\label{eq:minimizeH}
		\end{equation}

	In summary, we took a set of conditions, described by equations (\ref{eq:strongOE}-\ref{eq:strongHeredity}), valid in
	the general AIT framework, and we have re-written them in terms of statistical entropy functions through equations
	(\ref{eq:H<H}, \ref{eq:unboundedH}) and (\ref{eq:minimizeH}). We finally observe that the condition that the
	probability distribution must be strictly ordered leads to:
		\begin{equation}
			H(X_n)<\log n	\quad.\nonumber
		\end{equation}
	Accordingly, the case of total randomness (fair coin toss) is removed.


	\subsection{Minimizing the differences between shared states}
		\label{sec:3.01}

		Condition (\ref{eq:minimizeH}) is difficult to handle directly. Nevertheless, it can be approached as follows: We
		first find a minimum by extremalising a given Kullback-Leibler divergence, and then we will prove that this solution
		indeed converges to the absolute minimum of $H(X_{n+1}|X_n)$.

		Let us define the distribution $\hat{p}_{n+1}$ as
			\[
				\hat{p}_{n+1}(k)\equiv p_{n+1}(k|k<n+1)=\frac{p_{n+1(k)}}{\sum_{i< n+1} p_{n+1}(i)}\quad. 
			\]
		$\hat{p}_{n+1}(k)$ is the probability that $k<n+1$ appears when we draw the random variable $X_{n+1}$ {\em excluding
		the outcomes $=n+1$}. Clearly, $\hat{p}_{n+1}$ and $p_n$ are defined over the set $\Omega_{n}$, whereas $p_{n+1}$ is
		defined over the set $\Omega_{n+1}$. Since the support sets for both $\hat{p}_{n+1}$ and $p_n$ are the same, one can
		use the {\em Kullback-Leibler divergence} (KL-divergence) defined as the relative entropy (or information gain)
		between $p_n$ and $\hat{p}_{n+1}$:
			\begin{eqnarray}
				D(p_n||\hat{p}_{n+1}) &=& \sum_{k\leq n}p_{n}(k)\log {p_n(k) \over \hat{p}_{n+1}(k)}\quad.\nonumber
			\end{eqnarray}
		Now we impose the condition of path dependence as a variational principle over the K-L divergence and then we write
		the following Langrangian which defines the evolution of our system:
			\begin{eqnarray}
				&&{\cal L}(\hat{p}_{n+1}(1), \dots, \hat{p}_{n+1}(n); \theta_{n+1})  =  \nonumber \\
				&&= D(p_n||\hat{p}_{n+1}) + \theta_{n+1}\left(\sum_{k\leq n}\hat{p}_{n+1}(k)-1\right)\quad. \nonumber
			\end{eqnarray}
		The minimization of this Lagrangian with respect to the variables upon which it depends imposes that:
			\begin{equation}
				\hat{p}_{n+1}=p_n\quad,\nonumber
			\end{equation}
		which implies that
			\begin{eqnarray}
				p_{n+1}(k)&=&\theta_{n+1} p_n(k)\quad \forall k\leq n, \nonumber\\
				p_{n+1}(n+1)&=&1-\theta_{n+1}\quad.
				\label{eq:ptheta}
			\end{eqnarray}
		By construction, $0<\theta_{n+1}<1$.  Equation (\ref{eq:ptheta}) imposes that the conditional probabilities between
		$X_{n}$ and $X_{n+1}$ read: 
			\begin{equation}
				\left\{ \begin{array}{ll}
									\mathbb{P}_n(X_{n+1}=i| X_n=k)=\delta_{ik}\theta_{n+1}, & \text{for } i \leq n; \\
									\mathbb{P}_n(X_{n+1}=n+1| X_n=k)=1-\theta_{n+1}, & \text{for } k\leq n. 
								\end{array}
				\right.\quad\nonumber
			\end{equation}
		This defines a channel structure that leads to:
			\begin{equation}
				H(X_{n+1}|X_n)= H(\theta_{n+1})\quad,
				\label{eq:H(0_n)}
			\end{equation}
		being $H(\theta_{n+1})$ the entropy of a Bernoulli process having parameter $\theta_{n+1}$, i.e:
			\[
				H(\theta_{n+1})=-\theta_{n+1}\log\theta_{n+1}-(1-\theta_{n+1})\log(1-\theta_{n+1})\quad. 
			\]
		In the appendix it is proven that
			\[
				H(\theta_{n+1})\to\min H(X_{n+1}|X_n)\quad.
			\]
		We thus have found the specific form of the conditional entropy governing the path dependency of the OEE system,
		imposed by equation (\ref{eq:minimizeH}).
	
		We finally remark some observations related to the flow of information between past and present states. First, we
		note that, from equation (\ref{eq:H(0_n)}), the relation between the entropies of $X_n$ and $X_{n+1}$ satisfies the
		following Fano's-like equality:
			\begin{eqnarray}
				H(X_{n+1})=\theta_{n+1} H(X_n) +H(\theta_{n+1})\quad. 
				\label{eq:HthetaH}
			\end{eqnarray}
		Finally, from the definition of mutual information between $X_n$ and $X_{n+1}$, one gets:
			\[
				I(X_{n+1}:X_n)=H(X_{n+1})-H(X_{n+1}|X_{n}) \quad,
			\]
		and from equations (\ref{eq:H(0_n)}) and (\ref{eq:HthetaH}) we arrive at the amount of information transmitted from
		the time step $n$ to $n+1$:
			\begin{equation}
				I(X_{n+1}:X_n)= \theta_{n+1} H(X_n)\quad. 
				\label{eq:Bound(X:Y)}
			\end{equation}
		This is a good estimate of the maximum possible information transmitted per evolutionary time step. Nevertheless,
		even in this case, we shall see that the statistical information transmitted along time in an open-ended system has
		to face a paradoxical behaviour: the total loss of any past history in the long run --see section \ref{sec:Loss}.

	\subsection{Zipf's Law: the footprint of OEE}
		\label{sec:3.03}

		As discussed at the beginning, a remarkably common feature of several systems known to exhibit  OEE is the presence
		of Zipf's law.  We will rely now on previous results \cite{ZipfUniversal, CorominasMurtraSole2011} to show that the
		solution to the problem discussed above is given precisely by Zipf's law. We first note that, thanks to equation
		(\ref{eq:ptheta}), the quotient between probabilities:
			\begin{eqnarray}
				{p_n(i+j) \over p_{n}(i)} = f(i, i+j)\quad,\nonumber
			\end{eqnarray}
		remains constant for all $n$ as soon as $p_n(i+j)>0$. In the appendix, following
		\cite{CorominasMurtraSole2011}, we provide the demonstration that, in a very general case, the solution of our
		problem lies in the range defined by: 
			\begin{equation}
				\left(\frac{i+1}{i}\right)^{(1-\delta)}>f_n(i, i+1)>\left(\frac{i+1}{i}\right)^{(1+\delta)}\quad. \nonumber
			\end{equation}
		It can be shown that $\delta\to 0$ if the size of the system is large enough. Therefore:
			\begin{equation}
				f_n(i,i+1)=\frac{p_{n}(i)}{p_{n}({i+1})}\approx \frac{i+1}{i}\quad, \nonumber
			\end{equation}
		which leads us to the scaling distribution:
			\begin{equation}
				p_{n}(i)\propto i^{-1}\quad.
			\end{equation}
		In other words, Zipf's law is the only asymptotic solution,  which immediately suggests a deep connection between the
		potential for open-ended evolutionary dynamics and the presence of this particular power law. Note that Zipf's law
		is a necessary footprint of OEE, not a sufficient one: other mechanisms might imprint the same distribution
		\cite{NewmanZipf}. We emphasize the remarkable property that this result is independent of the particular way the
		evolving system satisfies the OEE conditions imposed by equations (\ref{eq:H<H}, \ref{eq:unboundedH}) and
		(\ref{eq:minimizeH}).

	\subsection{The loss of information paradox in OEE}
		\label{sec:Loss}

		\begin{figure}
			
			\begin{center}
				\includegraphics[width= 8.0 cm]{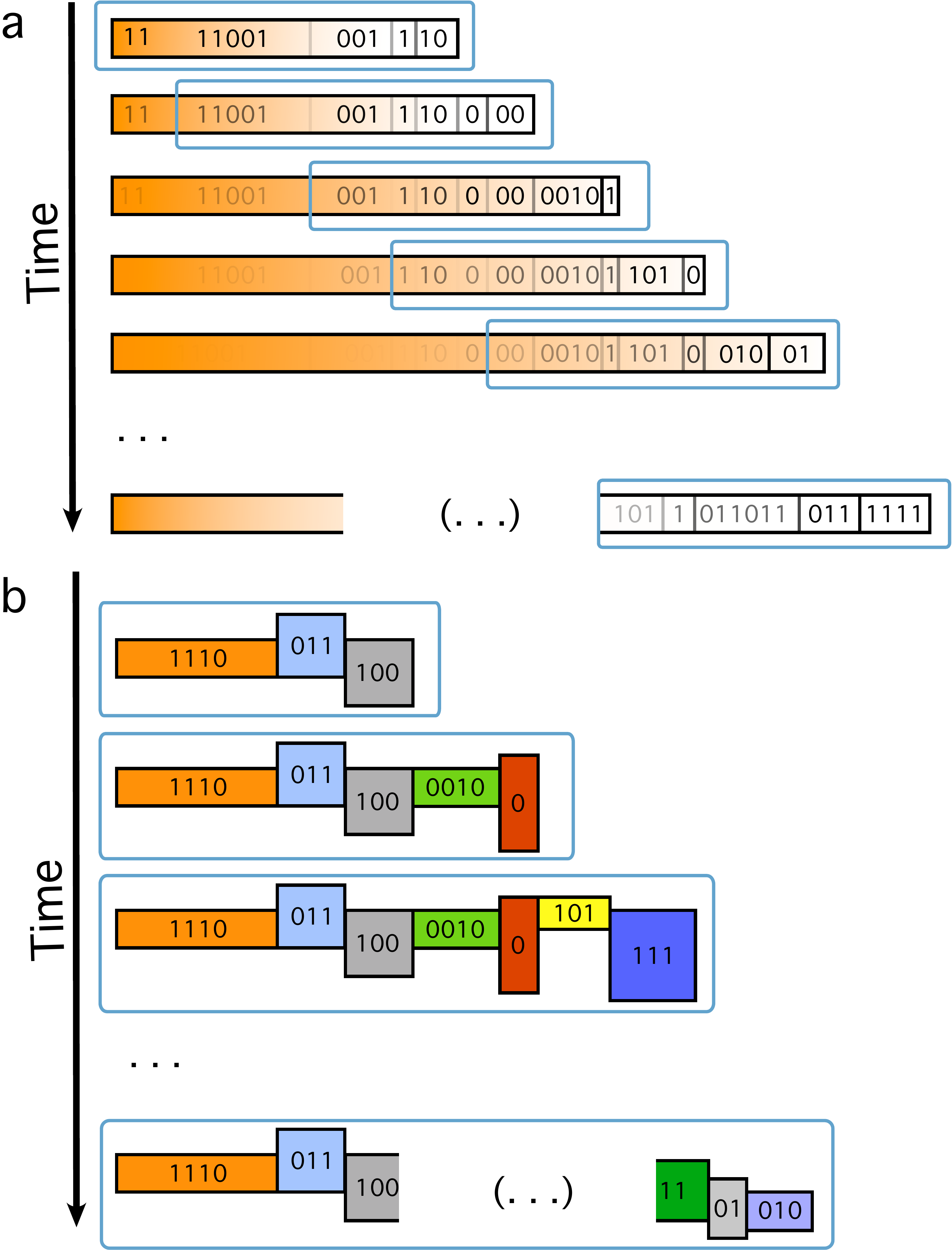}
			\end{center}

			\caption{The paradox of information loss: a) A statistical description of the system displays OEE. This means 
			that, through time, the entropy of the ensemble grows without bounds. The consequence of that is that the
			information about the past history is totally erased as time goes by. Only a fraction of the history closer to the
			present survives. Therefore, there is no conservation of the information -- see section \ref{sec:Loss}. b) If our
			information is encoded in a non-statistical way, such as bit strings, it can be preserved. The historical
			information survives through evolutionary stages, even if the system displays OEE.}
			\label{Fig:GrowingComplexity}
		\end{figure}

	The above description of the evolution of open-ended statistical ensembles leads to an unexpected result: Statistical
	systems displaying OEE loose any information of the past after a large period of complexity growing. Indeed, in spite
	information is conserved in a huge fraction step by step, it is not conserved at all if we compare large periods of
	evolution. Therefore, the capacity to generate an ensemble encoding an unbounded amount of information through
	evolution results in a total erasure of the past, even if a strong path dependency principle is at work.

	To see what happens with information along the evolutionary process in the limit of large $n$'s, we first rewrite
	mutual information between $X_n$ and $X_m$, $m<n$ as follows:
		\[
			I(X_m:X_n)=\sum_{i\leq n}p_n(i)\sum_{k\leq m}\mathbb{P}(k|i)\log\frac{\mathbb{P}(k|i)}{p_m(k)}\quad,
		\]
	where, in this case,  $\mathbb{P}(k|i)\equiv \mathbb{P}(X_m=k|X_n=i)$. Then we define the following constant $C_m$:
		\begin{equation} 
			C_m=\prod_{2\leq k\leq m}(\theta_k)^{-1}\quad, \nonumber
		\end{equation} 
	where the $\theta_k$'s are the ones arising from equation (\ref{eq:ptheta}).
	From here, one can prove --see appendix-- that: 
		\[
			p_m(1)=\frac{1}{C_m}\quad,
		\]
	Now, observe that we can generalize equation (\ref{eq:Bound(X:Y)}) as follows:
		\[
			I(X_m:X_n)\leq \theta_{m+1}\cdot . . . \cdot \theta_nH(X_m)\quad.
		\]
	This allows us to obtain the following chain of inequalities:
		\begin{eqnarray}
			I(X_m:X_n)&\leq & \prod_{m<i\leq n}\theta_i H(X_n)\nonumber\\
			&=&\frac{1}{C_n}\prod_{2\leq k\leq m}\theta_k^{-1}H(X_m)\nonumber\\
			&=&\frac{C_m}{C_n}H(X_m)\quad. 
			\label{eq:cn/cm}
		\end{eqnarray}
	The above inequalities have an interesting consequence. Indeed, from equation (\ref{eq:cn/cm}), if
	$C_n\to \infty$, then
		\begin{equation}
			\lim_{n\to \infty}I(X_m:X_n)\leq \lim_{n\to\infty}\frac{C_m}{C_n}H(X_m)=0\quad.
			\label{eq:lossInfo}
		\end{equation}
	In the appendix it is proven that, in OEE statistical systems, indeed we have that $C_n\to \infty$. Thus
	$I(X_m:X_n)\to 0$: {\em no statistical information is conserved in open-ended systems in the long term}.


	\subsection{Solving the paradox: Algorithmic information can be maintained}

		We have shown above that {\em statistical information} cannot be maintained through arbitrarily long evolutionary
		paths if the evolution is open-ended. The emphasis is on the word {\em statistical}. As we shall see, using a rather
		informal reasoning, other types of information based on the general setting of AIT can be maintained. Let $\sigma_n$
		be a description, in bits, of an object at time $n$ and $\sigma_N$ its description at time $N>n$. Let us assume that
		$\sigma_N$, in its most compressed form, can only be written as a concatenation of two descriptions, to be indicated
		with symbol ``$\oplus$'':
			\[
				\sigma_N=\sigma_n\oplus\sigma_{N-n} \quad.
			\]
		Now assume that $K(\sigma_N)=\mu N$, $K(\sigma_n)=\mu n$ and $K(\sigma_{N-n})=\mu(N-n)$, with $0<\mu<1$. If $\pi_n$
		is the minimal program that prints $\sigma_n$ and $\pi_{N-n}$ is the minimal program that prints $\sigma_{N-n}$.
		Then, there is a program $\pi_N$ defined as
			\[
				\pi_N=\pi_{n}\oplus\pi_{N-n} \quad,
			\]
		such that, when applied to a Universal Turing machine, gives $\sigma_N$, i.e., ${\cal T}_u(\pi_N)=\sigma_N$. If we
		already know $\pi_n$, it is clear that
			\[
				K(\sigma_N|\sigma_n)= K(\sigma_{N-n})+{\cal O}(1) \quad.
			\]
		We observe that, under the assumptions we made, 
			\[
				|K(\sigma_N)-K(\sigma_n)|=K(\sigma_{N-n}) \quad,
			\]
		so $K(\sigma_N|\sigma_n)\approx |K(\sigma_N)-K(\sigma_n)|$ close to the bound provided by Zurek in \cite{Zurek1989},
		already used in equation (\ref{eq:boundZurek}). As we shall see, the immediate consequence of that is that the
		algorithmic mutual information between $\sigma_N$ and $\sigma_n$ {\em does not depend on $N$}. Let
		$I(\sigma_N:\sigma_n)$ be the algorithmic mutual information between $\sigma_N$ and $\sigma_n$:
			\[
				I(\sigma_N:\sigma_n)=K(\sigma_N)-K(\sigma_N|\sigma_n) \quad.
			\]
		Then, one has that:
			\begin{eqnarray}
				I(\sigma_N:\sigma_n)&=&K(\sigma_N)-K(\sigma_N|\sigma_n)\nonumber\\
				&\approx&K(\sigma_N)-K(\sigma_{N-m})\nonumber\\
				&\approx&K(\sigma_n)\quad,\nonumber 
			\end{eqnarray}
		we thus have
			\begin{equation}
				\lim_{N\to\infty}I(\sigma_N:\sigma_n)\approx K(\sigma_n) \quad. 
			\end{equation}
		Within the algorithmic information theory framework this implies that information of previous stages of the
		evolution can be maintained.

		The result reported above has an important consequence: In an OEE system in which information is maintained, the
		{\em information} is encoded by generative rules that cannot be captured by simple statistical models. Therefore,
		Shannon information theory is of little use to understand the persistence of the memory of past states in OEE
		systems.

\section{Discussion}
	\label{sec:4}

	In this paper we have considered a new approach to a key problem within complex systems theory and evolution, namely,
	the conditions  for {\em open-ended evolution} and its consequences.  We provided a general formalization of the
	problem through a small set of {\em postulates} summarized by equations (\ref{eq:K<K1}-\ref{eq:strongHeredity}) based
	on the framework of algorithmic information theory.  Despite the high degree of abstraction --which allows us to
	extract very general results-- important specific conclusions can be drawn: (i) In statistically describable systems,
	Zipf's law is the expected outcome of OEE. (ii) OEE systems have to face the {\em statistical information loss
	paradox}: Shannon information between different stages of the process tends to zero, and all information of the past
	is lost in the limit of large time periods. (iii) This paradoxical situation is solved when considering 
	non-statistical forms of information, and we provided an example where algorithmic information between arbitrary time
	steps is maintained. This result, however, does not invalidate previous approaches of statistical information theory
	concerning the study of {\em flows} of information within the system \cite{Tkacik:2016}, since our result refers to
	the structural complexity of the evolving entity. It is important to stress that {\em information} may unfold in
	several meanings or formal frameworks when talking about evolving systems. Moreover, further explorations should
	inquiry on the role of information flows in keeping and promoting the increase of structural complexity of evolving
	systems. In addition, it is worth to emphasize that, at the current level of development, our framework might fail to
	incorporate some processes, such as exaptation or abiotic external drives, that are not fully algorithmic but
	identified as key actors in evolutionary systems \cite{deVladarSzathmary2017}.  All these issues are relevant in order
	to understand and eventually {\em build} OEE systems, as it is the case within the context of artificial life
	\cite{TaylorWiser2016, Paleontology} by considering the possibility of building a system able to evolve under
	artificial conditions and maintain a constant source of {\em creativity} \cite{Investigations,Clune2016}.

	Since Zipf's law is the outcome of a statistical interpretation of the OEE postulates given in equations
	(\ref{eq:K<K1}-\ref{eq:strongHeredity}), one may be tempted to conclude that information is not conserved in those
	systems exhibiting Zipf's law in its statistical patterns. Instead, in line with the previous paragraph, it is
	important to stress that the statistical ensemble description can be just a partial picture of the system, and that
	other mechanisms of information prevalence, not necessarily statistic, are at work. Therefore, if our system exhibits
	Zipf's law and we have evidence of information conservation, the statistical pattern may be interpreted as the
	projection of other types of non-statistical information to the statistical observables.

	Biological systems exhibit marked potential capacity for  OEE resulting from their potential for growing and exploring
	new states and achieving novel functionalities. This open-endedness pervades the apparently unbounded exploration of
	the space of the possible. The two biological systems cited in the introduction, namely human language and the protein
	universe, share  the presence of an underlying grammar, which both enhances and constrains their combinatorial
	potential.  Analogously, the example provided by models of evolution through gene duplication or tinkering revealed
	that scaling laws  and other properties displayed by protein networks emerge from the amplification phenomena
	introduced by growth through copy-and-paste dynamics \cite{proteome1,proteome2,Dorogo2003}. One way  of doing this is
	provided by the tinkered nature of evolutionary change,  where systems evolve by means of extensive reuse of previous
	parts \cite{Jacob1977,Sole2002,Carlos2005}. This mechanism fully matches our assumptions:  generative rules that allow
	a way of expanding the state space, while the redundant nature of the process allows keeping most of the previous
	structures.

	We reserve a final word for a general comment on the role of OEE in the theory of biology.  Postulates described by
	equations (\ref{eq:K<K1}-\ref{eq:strongHeredity}) explicitly relate OEE to unpredictability. This, according to
	classic results like the {\em No free lunch theorem} \cite{Wolpert:1997}, puts a question mark on the possibility of a
	theory of evolution in the sense of classical physics. This issue, discussed also in \cite{Day2012}, may exclude the
	possibility of a predictive theory in terms of the explicit evolutionary innovations that will eventually emerge.
	Nevertheless, in this paper we prove that this is not an all-or-nothing situation: Interestingly, the postulates of
	OEE, which rule out the existence of a predictive theory, are precisely the conditions that allow us to identify one
	of the possible statistical regularities --Zipf's law-- governing such systems and thereby make {\em predictions} and,
	eventually, propose {\em physical} principles for them, adding a new, unexpected ingredient to the debate on
	predictability and evolution \cite{Wim:2016}. According to that, these principles would predict the statistical
	observables, but not the specific events that they represent.

\vspace{0.2 cm}

{\bf Author contributions}
BC-M, LFS and RS contributed to the idea, development, the computation of mathematical derivations and to the writing of the manuscript.

{\bf Acknowledgments}

\vspace{0.1 cm}

	We thank Jordi Pi\~nero, Sergi Valverde, Jordi Fortuny, Kepa Ruiz-Mirazo and Carlos Rodr\'iguez-Caso and the members
	of the  Complex Systems Lab for useful discussions. BC-M wants to thank Stefan Thurner, Rudolf Hanel, Peter Klimek,
	Vittorio Loretto and Vito DP Servedio for useful comments on previous versions of the manuscript. This work has been
	supported by the Bot\'in Foundation, by Banco Santander
	through its Santander Universities Global Division, a MINECO FIS2015-67616 fellowship, the Secretaria  d'Universitats
	i Recerca del Departament d'Economia i Coneixement de la Generalitat de  Catalunya (RS and LS) and the Santa Fe Institute (RS).

\newpage

\appendix
\section{Every unbounded OEE process in the general sense contains an unbounded OEE process in the strong sense}
	\label{app:generalStrongOEE}

	As introduced in section IIIB, a dynamical system whose description at time step $t$ is $\sigma_t$ is the result of a
	process which history is recorded by $\Sigma(t) \equiv \{\sigma_1, \dots, \sigma_t\}$. This is the collection of the
	description of our dynamical system at each and every time step until $t$. Note that $\Sigma(t) \subset \Sigma(t+1)$
	for every $t$. Also, note that a given evolutionary history might contain partial evolutionary histories. Imagine, for
	example, that we fail to record every other instantaneous description. This is likely in empirical setups: we might be
	able to record a system only once a minute, or once a day.

	More rigorously, take a sorted, infinite subset of the natural numbers $T \equiv \{t_1, t_2, t_3, \dots \} \subset
	\mathbb{N}$. Note that each $t_\tau$ is an integer and that not necessarily all integers appear in $T$, but that this
	set itself can be labeled by an index $\tau$ which runs over all the natural numbers. At each $\tau$, the finite set
	$T_\tau \equiv \{t_1, \dots, t_\tau \} \subset T$ selects a subset $\Sigma'(\tau) \equiv \{\sigma_{t_1}, \dots,
	\sigma_{t_\tau}\}$ of the original history at time $t_\tau$, this is: $\Sigma'(\tau) \subset \Sigma(t_\tau)$. Also,
	$\Sigma'(\tau) \subset \Sigma'(\tau+1)$. We say that the succession of $\Sigma'(\tau)$ for all $\tau \in \mathbb{N}$
	is a partial history of the process under research. We also say that the original history (given by the succession of
	all $\Sigma(t)$ for all $t \in \mathbb{N}$) contains this partial history. \\

	With these definitions it is possible to prove that every unbounded OEE process in the general sense must contain an
	unbounded OEE process in the strong sense (as illustrated in figure \ref{fig:01}). Let us suppose, indeed, that our
	$\Sigma(t)$ obeys equations (6) and (7) of the main text. At the same time, let us also assume that, among all partial
	histories of this process, there is not a single one that obeys equation (9) of the main text -- this is, that our
	unbounded OEE process in the general sense does not contain any open-ended partial history in the strong sense. This
	second assumption will bring us to a contradiction.
	\begin{figure}
		\begin{center} 
			\includegraphics[width=0.85\columnwidth]{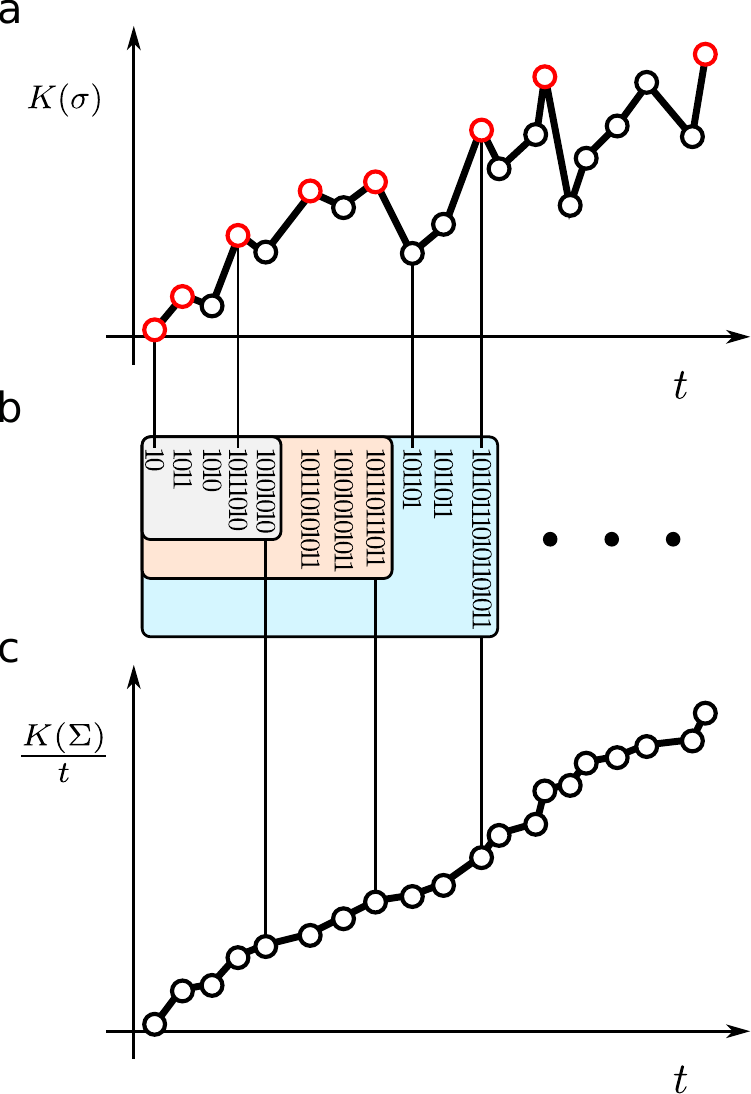}
		\end{center} 
    \caption{{\bf OEE in the general versus the strong sense}. a) The description of a system over time changes,
not necessarily yielding systems of monotonously growing complexity.  b)The evolutionary history of a system at a
given time ($\Sigma(t)$) consists of the collection of all previous description of the system up to time $t$. c) If
we deal with OEE in the general sense, the complexity (properly normalized) of the history of a system increases
monotonously over time. As we prove in the text, if such a system complies with all axioms for general OEE, it must
contain an OEE sub-process in the strong sense, marked with red circles in panel  a). }
		\label{fig:01}   
	\end{figure}

	The fact that there is not any partial history obeying equation (9) of the main text means that, for whichever partial
	history that we choose, there is always a finite value $\mu$ such that $K(\sigma_{t_{\mu}}) > K(\sigma_{t_\tau})$ for
	all $\tau > \mu$. This implies that the description of our system reaches a maximum $K^+$ at some time $m \leq t_\mu$
	with $K^+ \equiv K(\sigma_{m})$, and that every description of the system afterwards has at most complexity $K^+$. The
	complexity of the process history at this time normalized by the number of steps is some finite number:
		\begin{eqnarray}
			\left<K(\Sigma(m))\right> = {K(\Sigma(m)) \over m} < \infty\quad. 
		\end{eqnarray}

	Let us now study the case in which as much complexity as possible is added after this time step. Note that
	$K(\sigma_{t>m})$ is {\em at most} $K^+$. To add as much complexity as possible to the process history, we would need
	to append incompressible bit strings. The number of incompressible strings of length $K^+$ is finite, but we can still
	add them in a patternless fashion, also making sure correlations are not introduced with the history prior to time
	$m$. Even if we manage to do so, the complexity of the process history (normalized by the number of time steps) at any
	time $t>m$ is bounded by:
		\begin{eqnarray}
			K(\Sigma(t>m)) < { m \left<K(\Sigma(m))\right> + (t-m)K^+ \over t}\quad. 
		\end{eqnarray}

	For large enough times, $K(\Sigma(t>m)) \rightarrow K^+$ asymptotically. By axiom 1 (Open-Endedness), this process
	would be a bounded open-ended one if $K^+ > \left<K(\Sigma(m))\right>$. This is: bounded open ended processes in the
	general sense do not necessarily contain an open-ended process in the strong sense. But the upper limit $K^+$ to the
	complexity of the process history implies that our process cannot obey axiom 2 (Unboundedness), which is a
	contradiction because we started out by assuming that our process is unbounded and open-ended in the general sense.
	Hence, every unbounded open-ended process in the general sense must contain at least a partial history which is 
	open-ended in the strong sense.

	We could proceed similarly to prove that at least one partial history must exist that is unbounded in the strong
	sense. If not, all partial histories must have a finite $N \in \mathbb{N}$ such that there is not any $t$ such that
	$K(\sigma_t) > N$. All of these bounds are finite, so there must be a maximum $K^+ = \max\{N_i\}$, where $N_i$ is the
	bound of the $i$-th partial history of the original process (which is a countable set). We can try to build the most
	complex such history and come to the conclusion that, {\em at most}, $K(\Sigma(t)) \rightarrow K^+$ as $t \rightarrow
	\infty$. But, again, we departed from the hypothesis that our process is unbounded open-ended, so this is a
	contradiction because $K^+$ sets an upper bound in complexity of the process history normalized by number of time
	steps. Hence every unbounded open-ended process in the general sense must contain an unbounded open-ended process in
	the strong sense.

\section{Conditions emergence of order in the probability distribution}
	\label{Sec:CondOrder}

Throughout the text we emphasised that the probability distribution is ordered. It is therefore crucial that such ordering is maintained. Here we state the conditions under which the emerging probability distribution is ordered.

The solution that satisfies the postulates of OEE given by equations (13--15) of ther main text assumes the existence of some sequence $\theta_2, . . .,\theta_n,...$ of positive real numbers by which:
\begin{eqnarray}
(\forall k\leq n)&& p_{n+1}(k)=\theta_{n+1}p_n(k);\nonumber\\
{\rm if}\;k=n+1;&&p_{n+1}(k)=1-\theta_{n+1} \quad.
\end{eqnarray}
This implies a successive process of rescaling of the probabilities, as long as the system grows in size. So, one has:
\begin{eqnarray}
p_n(1)&=&\theta_2\cdot\theta_3\cdot . . . \cdot\theta_n=\prod_{k\leq n}\theta_k\nonumber\\
p_n(2)&=&(1-\theta_2)\prod_{2<k\leq n}\theta_k\nonumber\\
. . .&=&. . .\nonumber\\
p_n(i)&=&(1-\theta_i)\prod_{i<k\leq n}\theta_k\nonumber\\
. . .&=& . . .\nonumber\\
p_n(n)&=&1-\theta_n\quad.\nonumber 
\end{eqnarray}
We observe that we can establish a recurrence relation between probabilities:
\begin{eqnarray}
p_n(1)&=&\prod_{1<k\leq n}\theta_k\nonumber\\
p_n(k)&=&a_kp_n(k-1)\quad,\nonumber 
\label{eq:pak}
\end{eqnarray}
with $a_k$ defined as:
\[
a_k\equiv \frac{(1-\theta_k)}{(1-\theta_{k-1})\theta_k} \quad.
\]
If $a_k=1$, then $(\forall k\leq n)\;p_n(k)=\frac{1}{n}$, and $H(X_n)=\log n$. It is easy to see that
\[
\theta_k=\frac{k-1}{k}\Rightarrow (\forall k\leq n)\;p_n(k)=\frac{1}{n} \quad,
\]
and, consistently, $1-\theta_k=\frac{1}{k}$.
Now let us suppose that the function $1-\theta_k$ is dominated by $\frac{1}{k}$, i.e., $1-\theta_k$ decays faster than $\frac{1}{k}$. Then,
\[
\frac{(1-\theta_k)}{(1-\theta_{k-1})}<\frac{k-1}{k}\quad{\rm and}\quad\theta_k>\frac{k-1}{k} \quad,
\]
so
\[
(\forall k\leq n)\quad a_k= \frac{(1-\theta_k)}{(1-\theta_{k-1})\theta_k}<1 \quad.
\]
The immediate consequence of the above result is that:
\begin{equation}
p_n(1)>p_n(2)> . . .>p_n(n) \quad.
\label{eq:eordering}
\end{equation}
We observe that if $H(X_n)<\log n$ then $\theta_k\neq \frac{k-1}{k}$. We impose that the solution taken is the one giving
\[
1>\theta_k> \frac{k-1}{k}\quad,
\]
such that equation (\ref{eq:eordering}) is satisfied.

\section{Minimisation of conditional entropy though the K-L divergence}
\label{appendix:KLDiv}

In this section we will prove that the solution provided by the minimisation of the K-L divergence converges to the absolute minimum of $H(X_{n+1}|X_n)$ in an OEE statistical system. This implies that, even we cannot prove that this is the absolute solution, we can prove that it is arbitrarily close to it.

Let us have the following relation between successive probability distributions:
\begin{eqnarray}
p_{n+1}(k)&=&\theta_{n+1} p_n(k)\quad \forall k\leq n\nonumber\\
p_{n+1}(n+1)&=&1-\theta_{n+1} \quad,
\end{eqnarray}
which is the solution of the minimisation of the K-L divergence as shown in section IVA of the main text.
This leads to an amount of noise:
\begin{equation}
	H(X_{n+1}|X_n)= H(\theta_{n+1})\quad,
	\label{eq:H(0)}
\end{equation}
being $H(\theta_n)$ the entropy of a Bernoulli process having parameter $\theta_n$, i.e:
	\[
		H(\theta_n)=-\theta_n\log\theta_n-(1-\theta_n)\log(1-\theta_n) \quad.
	\]
Let $\min H(X_{n+1}|H(X_n))\leq H(\theta_n)$ be the absolute minimum of $H(X_{n+1}|X_n)$ under the conditions of OEE described in equations (13--15) of the main text. We will show that $(\forall \epsilon>0)$ $\exists M$ for which, for any $N>M$:
\[
\left|\min H(X_{n+1}|H(X_n))-H(\theta_N)\right|<\epsilon \quad.
\]
Indeed, let us suppose that our system is open-ended. This implies that $1>\theta_n\geq (n-1)/n$ --see section \ref{Sec:CondOrder} of this appendix. So, knowing that, by definition $H(\theta_n)=H(1-\theta_n)$ we have that for any $\epsilon'>0$ $\exists M$ such that, for any $n>M$ $1-\theta_n<\epsilon'$. This implies that, for any $\epsilon>0$, $\exists M$ such that, for any $n>M$:
\[
H(\theta_n)<\epsilon \quad.
\]
Since $H(X_{n+1})>H(X_n)$ by the postulates of OEE, then $\min H(X_{n+1}|H(X_n))>0$. In addition, we have proven that $H(\theta_n)<\epsilon$. Therefore, since, by assumption $\min H(X_{n+1}|H(X_n))\leq H(\theta_n)$, we demonstrated, taking the $\epsilon$ above defined, that:
$(\forall \epsilon>0)$ $\exists M$ for which, for any $n>M$:
\[
\left|\min H(X_{n+1}|H(X_n))-H(\theta_n)\right|<\epsilon \quad.
\]
Therefore, $H(\theta_n)$ converges asymptotically to the absolute minimum.


\section{Derivation of Zipf's law from entropy constraints}

\label{App:Zipf_proof}

Assume  that the unboundedness condition is given as follows: there exists a unique $ \mu\in  (0,1)$  such  that
$(\forall \epsilon >0)(\exists N):(\forall n>N)$:
\begin{equation}
\left| \frac{H(X_n)}{\log n}-\mu\right|<\epsilon\quad.
\label{Unboundedness}
\end{equation}
Now we  want to find the  asymptotic behavior of  $p_n, \;n\to \infty$
under  the  above  justified  conditions  given by equation (\ref{Unboundedness})  of this appendix and
(16) of the main text.   The  key   feature  is  that  the
following quotient:
\begin{equation}
(\forall k+j\leq n)\;\;f(k,k+j)=\frac{p_n(k+j)}{p_n({k})}\quad,
\label{f(m,m)}
\end{equation}
does not depend on $n$.  Therefore, along the evolutionary process, as
soon  as 
\[
p_n(k),p_n({k+j})>0\quad,
\]
$f(k,k+j)$ remains  invariant. 

Now suppose that wehave $p'_n\sim i^{-\gamma}$. The explicit form of its (normalized) entropy is:
\begin{equation}
\frac{H(X'_n)}{\log n}=\frac{1}{\log n}\left(\frac{\gamma}{Z_{\gamma}}\sum_{i\leq n} \frac{\log i}{i^{\gamma}}+\log Z_{\gamma}\right)\quad.
\label{H(Y)}
\end{equation} 
where $Z_{\gamma}$ is the normalization constant. From the above expression,  we  find that,  if  $(\forall \delta >0,\;n>m)(\exists N)$ such that:
\begin{equation}
(\forall
  m>N)\;\;\;f(m,m+1)<\left(\frac{m}{m+1}\right)^{1+\delta}\quad,\nonumber
\end{equation}
then
$(\exists   C<\infty  \in   \mathbb{R}^+)\;{\rm  such\;that}\;(\forall n)(H(X_n)<C)$, leading to 
\begin{eqnarray}
\lim_{n\to \infty}\frac{H(X_n)}{\log n}=0\quad,\nonumber
\end{eqnarray}
which  contradicts the  assumptions of  the problem,  depicted  by equation
(\ref{Unboundedness}). Therefore, during
the growth process,
\begin{equation}
f(m,m+1)>\left(\frac{m}{m+1}\right)^{(1+\delta)}\quad,
\label{>1+d}
\end{equation}
with  $\delta$  arbitrarily  small,  provided that  $n$  can  increase
unboundedly.  
%
%
Furthermore, we observe that, if $(\forall \delta >0,\;n>m)(\exists
N)$ such that
\begin{equation}
(\forall
  m>N)\;\;f(m,m+1)>\left(\frac{m}{m+1}\right)^{(1-\delta)}\quad,\nonumber
  \label{Dominated}
\end{equation}
then, from equation (\ref{H(Y)}), one finds that:
\begin{eqnarray}
\lim_{n\to \infty}\frac{H(X_n)}{\log n}=1\quad,\nonumber
\end{eqnarray}
again in  contradiction to equation (\ref{Unboundedness}),  except in the
extreme, pathological case where $\mu=1$,
which has been ruled out by assumption. 
 Accordingly, 
 \begin{equation}
f(m,m+1)<\left(\frac{m}{m+1}\right)^{(1-\delta)}\quad.
\label{<1-d}
\end{equation}
Combining equation (\ref{>1+d})  and (\ref{<1-d}), we have shown that
the  asymptotic  solution  is   bounded  by  the  following  chain  of
inequalities:
\begin{equation}
\left(\frac{m}{m+1}\right)^{(1+\delta)}<f(m,m+1)<\left(\frac{m}{m+1}\right)^{(1-\delta)}\quad.\nonumber
\end{equation}
The crucial step is that it, if $\mu\in(0,1)$, using the fact that equation (\ref{H(Y)}) defines a continuous, smooth function in terms of $\gamma$, one can conclude that for $n\to\infty$,
\[
\delta\to 0\quad.
\]
This implies, in turn, that, for $n\gg 1$:
\[
f(m,m+1)\approx \frac{m}{m+1}\quad,
\]
and, from  the definition  of $f$ provided  in equation  (\ref{f(m,m)}), we
conclude that:
\begin{equation}
p_{n}(k)\propto \frac{1}{k}\quad,\nonumber
\end{equation}
leading  us to  Zipf's  law  as the  unique  asymptotic solution.

\section{Divergence of the normalization constant}

\label{Appendix:Divergence}

Given the expression of $p_n(1)$ obtained above:
 \[ 
 p_n(1)=\prod_{2<k\leq n}(\theta_k)  \quad,
 \] 
one can define the normalisation constant
\begin{equation} 
C_n=\prod_{2\leq k\leq n}(\theta_k)^{-1}  \quad.
\label{eq:norm} 
\end{equation} 
We observe that we can rewrite the probability distribution 
 $p_n$ in the following form: 
 \[ 
 \frac{1}{C_n}, \frac{1-\theta_2}{C_n}, \frac{1-\theta_3}{\theta_2C_n},. . .  
 \] 
 following the above series, it is not difficult to see
that, $(\forall i)(1<i\leq n)$: 
\[ 
p_n(i)=\frac{1-\theta_i}{C_n}\prod_{2\leq k<i}(\theta_k)^{-1}\quad. 
\]
Now we connected the parameters related to the increase of the entropy during the evolutionary path and the normalisation constant of the distribution. This normalisation constant will be the key of our argument. Indeed, thanks to the properties of the Riemann $\zeta$ function it is known that, if $(\exists \epsilon >0):(\exists m):(\forall k>m)$ 
\[
\left(\frac{p_{n}(k+1)}{p_{n}(k)}\right)<\left(\frac{k}{k+1}\right)^{-(1+\epsilon)}\quad,
\]
(i.e. the probability distribution $p_n$ is {\em dominated} by the probability distribution $q_n(i)\propto i^{-1-\epsilon}$), then
$(\exists N\in \mathbb{N}):(N>C_{\infty})$, being
\[
C_{\infty}=\lim_{n\to\infty}C_n\quad.
\]
This means that, 
\[
(\forall n)\;p_n(1)\geq \frac{1}{C_{\infty}}>0\quad.
\]
However, in the case of $p_n$ not being dominated by any $q_n(i)\propto i^{-(1+\epsilon)}$, things go different, since
\[
\lim_{n\to\infty} C_n=\infty\quad.
\]
The direct consequence for the above consideration is that $(\forall \epsilon >0)(\exists M)$ such that, if $n>M$, then 
\[
p_n(1)<\epsilon\quad.
\]
The presence of an upper bound in the Shannon entropy is directly related to the divergence or convergence of $C_n$. If the probability distribution is not dominated by $q_n(i)\propto i^{-(1+\epsilon)}$ for any $\epsilon>0$, then the entropy, as well as $C_n$, diverges. On the contrary, if $p_n$ is dominated by $q_n(i)\propto i^{-(1+\epsilon)}$ for any $\epsilon>0$ the entropy converges and so does $C_n$. In formal terms:
\begin{equation}
\left(\lim_{n\to \infty}C_n=\infty\right)\Leftrightarrow \left(\lim_{n\to \infty}H(X_n)=\infty\right)\quad.
\label{eq:DoubleImplication}
\end{equation}
Zipf's law is thus at the twilight zone between bounded and unbounded complexity. Accordingly, for a OEE system under conditions described by equations (13--15), thanks to the bound on mutual information between an arbitrary past state $m$ and the current one $n$ given by equation (21) of the main text, one concludes that:
\[
\lim_{n\to \infty}I(X_m:X_n)\leq \lim_{n\to\infty}\frac{C_m}{C_n}H(X_m)=0\quad,
\]
in words, that all past information is lost.


\end{document}